\documentstyle[aps,pre]{revtex}

\input epsfig.sty

\title{Fast Monte Carlo algorithm for supercooled soft spheres}

\author{Tom\'as S.~Grigera and Giorgio Parisi}

\address{Dipartimento di Fisica, Universit\'a di Roma La Sapienza and
INFN Sezione di Roma La Sapienza,\\
Piazzale A. Moro 2, 00185 Roma, Italy}

\date{January 16, 2001}

\begin{document}

\draft

\wideabs{

\maketitle

\begin{abstract}

We present a nonlocal Monte Carlo algorithm with particle swaps that
greatly accelerates thermalization of soft sphere binary mixtures in
the glassy region. Our first results show that thermalization of
systems of hundreds of particles is achievable, and find behavior
compatible with a thermodynamic glass transition.

\end{abstract}


}


The dynamics and thermodynamics of supercooled liquids and glasses
have received much attention for some time, but are still an open
problem\cite{angell}. There has been theoretical progress regarding
equilibrium dynamics \cite{gotze} and thermodynamics (where older
ideas \cite{gibbs,adam} have been recently developed
\cite{mohanty,stillinger,mezardparisi,theoryss,theorylj,mezard}) as
well as non-equilibrium dynamics (mainly based on analogies with spin
glass behavior\cite{bouchaud}). Many issues are open \cite{theorsit},
such as the description of the dynamics of finite-range systems, or
the existence of a thermodynamic glass transition, which is not
accepted by all researchers \cite{krauth}.

Given the great complexity of an analytical treatment, here as in
other branches of science computer simulations play a major role
\cite{kobrev}.  However, simulations face difficulties, particularly
in the study of equilibrium properties\cite{kobrev}. Well-known
methods such as Molecular Dynamics (MD) and traditional Monte Carlo
(MC), with dynamics that closely resembles that of real systems,
suffer from a slow-down at temperatures approaching the glass
transition $T_g$. Just as a glass well below $T_g$ remains out of
equilibrium in the laboratory, model systems require prohibitively
long CPU time to equilibrate, even for a handful of particles.

Thus in recent years several algorithms have been proposed endowed
with a dynamics different from real systems, one that would allow them
to more easily jump across the free energy barriers that slow down MD
or MC. We can mention replica MC \cite{swendsen}, multicanonical MC
\cite{berg}, entropic sampling \cite{lee}, parallel tempering (PT)
\cite{marinari}, expanded ensembles MC \cite{lyubartsev}, and the
pivot cluster algorithm \cite{dress,malherbe}. Among these, parallel
tempering is very convenient because of its straightforward
implementation and ease of parallelization, and because it can be
applied to both MD and MC. It has been used successfully for spin
glasses for several years, and recently applied to structural glasses
\cite{coluzzi98,kob}. However, even with these improvements,
equilibration of deeply supercooled liquids requires a huge amount of
CPU time for systems of $N \approx 50$ or more particles.

In this paper we present a Monte Carlo algorithm with a simple but
nonlocal dynamics which is suited for the study of the soft sphere
binary mixture, a model structural glass former widely studied in the
past \cite{bernu,barrat,yip,parisi97}. This algorithm outperforms PT for
soft spheres. It allows to equilibrate small systems with modest CPU
time, and gives hope of equilibrating deeply supercooled systems of
hundreds of particles.

The soft spheres binary mixture is, with appropriately chosen
parameters, a fluid in which crystallization is strongly inhibited
\cite{bernu}. Half of the particles are of type $A$ and half of type
$B$, with radii $\sigma_A$ and $\sigma_B$ respectively and unit
mass. The potential energy is
\begin{equation}
{\cal V} = \sum_{i\neq j}^N \left( \sigma_i + \sigma_j \over |\mbox{\bf r}_i -
\mbox{\bf r}_j | \right)^{12}.
\end{equation}
The radii are determined by fixing their ratio to be
$\sigma_B/\sigma_A=1.2$ and by setting the effective diameter to unity,
which for equal number of $A$ and $B$ amounts to the condition
\begin{equation}
(2\sigma_A)^3 + 2(\sigma_a+\sigma_b)^3+(2\sigma_B)^3 = 4 l_0^3,
\end{equation}
where $l_0$ is the unit of length. Under these conditions, all excess
thermodynamic properties depend only on $\Gamma=\rho T^{-1/4}$, with
$\rho=N/V$ the density and $T$ the temperature. We have chosen
Boltzmann's constant $k_B=1$, and we restrict ourselves
to the case $\rho=1$. The glass transition is known to happen at
$\Gamma_c=1.45$ \cite{barrat}. Note that with this choices, the energy
and temperature are dimensionless.

We simulate this model with an algorithm that combines standard Monte
Carlo moves with nonlocal moves (particle swaps) \cite{pastore}. A
step of the algorithm can be described as follows:

\begin{itemize}

\item Choose a random particle $i$.

\item Draw 8 random numbers $r_1,\ldots,r_8$ uniform in $[0,1]$. 

\item If $r_1<1-p_S$ then
\begin{itemize}
\item Generate a standard MC trial move (i.e.\ shift particle by
 $\Delta \mbox{\bf r} = (2 r_2-1,2 r_3-1,2 r_4-1) \Delta r$)
\end{itemize}
else
\begin{itemize}
\item Choose a random particle $j$ of a different type.
\item Swap particle positions $\mbox{\bf r}_i \leftrightarrow \mbox{\bf
r}_j$.
\item Shift both particles as in the standard move ($\Delta \mbox{\bf
r}_i = (2 r_2-1,2 r_3-1,2 r_4-1) \Delta r$, $\Delta \mbox{\bf r}_j =
(2 r_5-1,2 r_6-1,2 r_7-1) \Delta r$).
\end{itemize}
 
\item Accept or reject new configuration according to the Metropolis
criterion, i.e.\ accept it if $ r_8 < \exp [ -\beta
(E_{\mbox{\scriptsize new}} - E_{\mbox{\scriptsize old}} ) ] $.

\item Repeat for all particles.

\end{itemize}

$\Delta r$ and $p_S$ are the algorithm parameters.  $\Delta r$ is
chosen at each temperature to achieve a shift acceptance ratio around
$0.5$, as is normally done in MC simulations. $p_S$ is the swap
attempt probability, which we choose equal to 0.1. Implementation of
this algorithm (which we shall call swap Monte Carlo (SW) algorithm)
is straightforward. Also, it should be obvious that SW satisfies
detailed balance.

Note that at high densities, even the swap of two neighboring
particles is a cooperative (hence slow) process. The particles must go
around each other, but they are hindered by the ``cage'' the rest of
the particles build around them. The idea behind the SW algorithm is
to accelerate these processes by making them non-cooperative.

We have run SW for systems of $N=34$ and $N=800$ particles in a cubic
box with periodic boundary conditions at several values of $\Gamma$ up
to $\Gamma=2$, deep in the glassy region. For the larger system, a
long-range cutoff at $r_c^2=3$ was imposed.

\begin{figure}
\epsfig{file=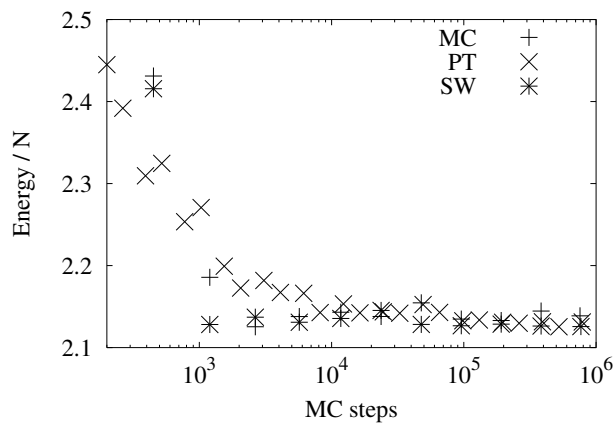, width=\columnwidth}
\caption{Comparison of the evolution the energy starting from a random
configuration for the three algorithms at $\Gamma=1.4$.}
\label{falgoc1}
\end{figure}

\begin{figure}
\epsfig{file=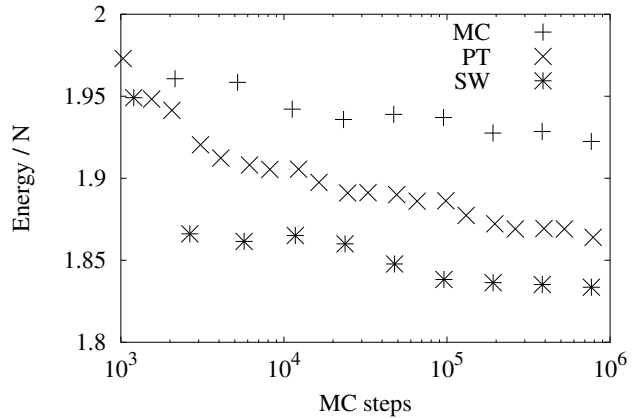, width=\columnwidth}
\caption{As fig.~\protect\ref{falgoc1} but for $\Gamma=1.8$.}
\label{falgoc2}
\end{figure}

We start by comparing the approach to equilibrium of a system of
$N=34$ particles simulated with standard MC, parallel tempering MC
and the SW algorithm. The PT data are from ref.~\cite{coluzzi98}
and were obtained in a PT run with 13 replicas with $\Gamma=1,
1.05,\ldots ,1.2, 1.3, \ldots ,2$. The evolution of the (potential)
energy per particle for each algorithm starting from a random
configuration is shown in figs.~\ref{falgoc1} and~\ref{falgoc2} for
temperatures above and below the transition. At $\Gamma=1.4$ (above
the transition), all algorithms equilibrate rapidly. At $\Gamma=1.8$,
however, it is clear that relaxation times have soared, and the three
algorithms behave differently. SW is seen to relax much faster than MC
and PT.

It may seem surprising that SW works at all, i.e.\ that the swaps are
ever accepted. But the radii of the spheres are not very different
from each other, and both species are present in equal amounts, so the
replacement of one species by the other is not always very
unfavorable. Indeed, the swap acceptance ratio is very small (at
$\Gamma=1.8$ it is about $10^{-4}$).  However, this means that a small
fraction of particles can escape the cage effect, with dramatic
consequences for the dynamics.

\begin{figure}
\epsfig{file=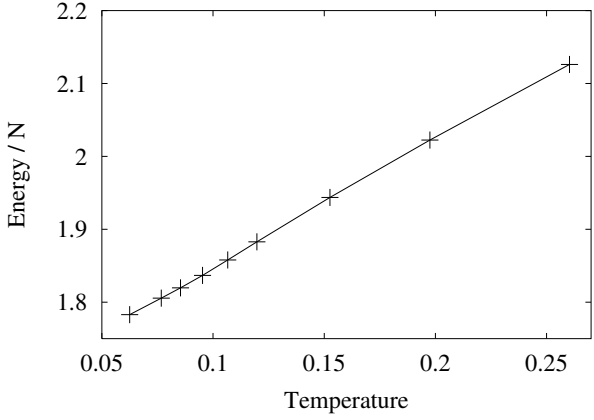, width=\columnwidth}
\caption{Average energy vs.\ temperature for $N=34$. Lines are only to
join neighboring points.}
\label{Egamma}
\end{figure}

\begin{figure}
\epsfig{file=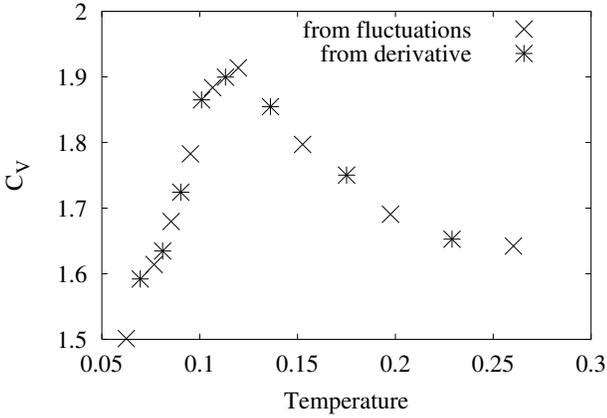, width=\columnwidth}
\caption{Specific heat vs.\ temperature for $N=34$. The values
obtained from the derivative of the energy and from energy
fluctuations fall on the same curve.}
\label{Cgamma}
\end{figure}

To check whether SW is capable of equilibrating the system we run
simulations with 2 different initial configurations at $\Gamma=1.4,
1.5,\ldots,2$. We find that after 1-2 million steps (about 80 minutes
of CPU time on an Alpha workstation), both configurations stabilize at
the same energy. The runs are then continued up to $10^7$ steps,
using the last $8\cdot 10^6$ steps to compute the energy and specific
heat. We check thermalization in two ways:
\begin{itemize}
\item We divide the last part of the run in blocks of 1 million steps
and verify that there is no systematic shift in the average energy of
these blocks.

\item We compute the specific heat in two different ways, through $C_V =
\partial \langle E \rangle /\partial T$ and through $C_V = (\langle E^2
\rangle -\langle E \rangle^2)/T^2$, and find agreement between the
two.
\end{itemize}
The energy and specific heat are shown in figs.~\ref{Egamma} and
\ref{Cgamma} as a function of temperature. The specific heat obtained
from fluctuations agrees with the value from the derivative of the
energy, showing that the system has thermalized. A broad but clear
peak around $T\approx 0.12$ is found \cite{fynewever}.

\begin{figure}
\epsfig{file=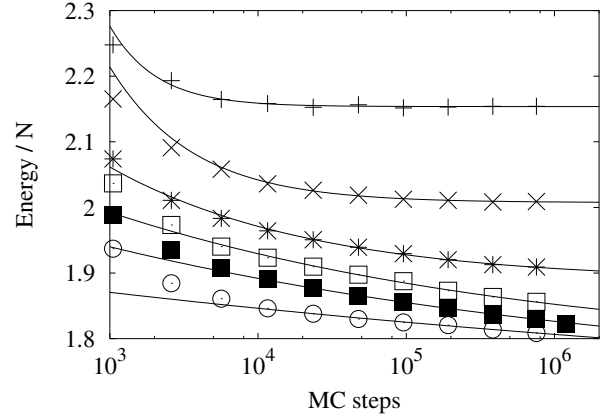, width=\columnwidth}
\caption{Energy vs.\ Monte Carlo time for $N=800$, averaged over 4
initial configurations. From top to bottom,
$\Gamma=1.4, 1.5, 1.6,$ $1.7, 1.8, 2$. Lines are power law fits.}
\label{evt800}
\end{figure}

\begin{figure}
\epsfig{file=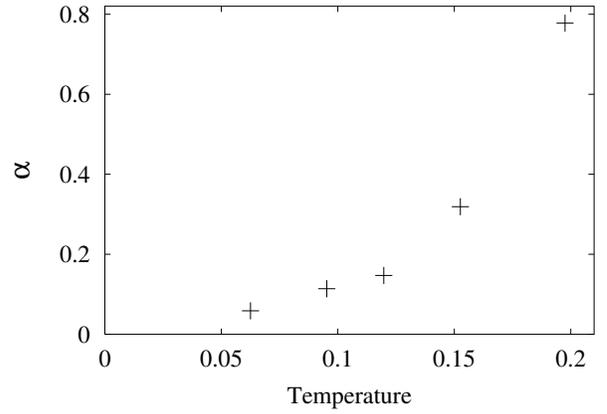, width=\columnwidth}
\caption{Exponent $\alpha$ of the energy relaxation vs.\ temperature.}
\label{alphaT}
\end{figure}

\begin{figure}
\epsfig{file=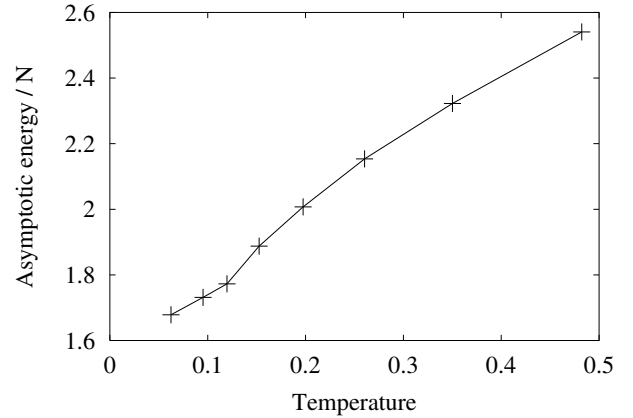, width=\columnwidth}
\caption{Asymptotic energy vs.\ temperature. Lines are only to join
neighboring points.}
\label{ET}
\end{figure}

\begin{figure}
\epsfig{file=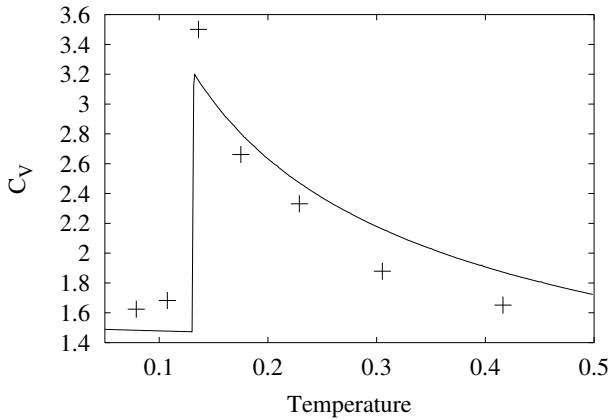, width=\columnwidth}
\caption{Specific heat vs.\ temperature for $N=800$ obtained from the
asymptotic value of the energy (points) with theoretical prediction
from ref.~\protect\cite{theoryss} (continuous curve).}
\label{cp800}
\end{figure}

Next we simulate a system of $N=800$ particles at $\Gamma=1.2, 1.3,
\ldots, 1.8, 2$. Fig.~\ref{evt800} show the energy relaxation
for runs of $10^6$ steps. The lowest temperatures clearly have not
reached equilibrium in this case.  The energy relaxation (after an
initial fast decay which corresponds to the equilibration time for the
liquid phase, namely $10^4$ steps) can be described by a power law
$E=E_0+t^{-\alpha}$, with $\alpha$ strongly temperature dependent
(fig.~\ref{alphaT}). A sharp change in slope of the $\alpha$ vs.\ $T$
curve is observed near $T\approx 0.15$. The asymptotic energy
(fig.~\ref{ET}) also shows singular behavior at this temperature, as
evidenced by the specific heat (fig.~\ref{cp800}) estimated from the
derivarive of the asymptotic energy.

Equilibrium simulations at low temperatures are relevant to the
question of the existence of a thermodynamic glass transition driven
by an entropy crisis \cite{gibbs,kauzmann}. In this respect it is to
note that the peak in the specific heat of the $N=34$ system
(fig.~\ref{Cgamma}) at $T\approx 0.12$ corresponds to $\Gamma\approx
1.7$, i.e.\ close to the predicted \cite{theoryss} critical
temperature $\Gamma_K\approx 1.65$ at which a thermodynamic transition
would take place and where the calculated specific heat shows a sharp
drop to a value of $3/2$ in the low temperature phase.

Remarkably, a similar but sharper behaviour is found for the $N=800$
system, though admittedly these values, resulting from extrapolation,
should be taken with care. In figure~\ref{cp800} we also show the
theoretical prediction for the specific heat (from~\cite{theoryss}).
There is a difference in the liquid phase, which is to be expected
since the hyper-netted chain approximation (HNC) used is known not to
perform very well at low temperatures. But the position of the jump, at
$\Gamma\approx 1.7$, is in good agreement with the theoretical
$\Gamma_K\approx 1.65$ and with the position of the peak for the
$N=34$ system.  The sharp change in the slope of the dynamic exponent
$\alpha$ vs.\ $T$ curve takes place at this same temperature.

In summary, we have presented a very efficient MC algorithm for
equilibration of a soft spheres binary mixture. We have managed to
equilibrate an $N=34$ system up to $\Gamma=2$ in roughly $2 \cdot
10^6$ steps (about 80 minutes of CPU time on an Alpha
workstation). Thus the SW algorithm puts thermalization of undercooled
systems of hundreds of particles within reach of modern CPUs. Our
first results with SW find behavior compatible with a thermodynamic
glass transition.

It would be very interesting to apply SW to other binary systems.
However, it is likely that it will need some modification to achieve a
reasonable swap acceptance ratio.

We thank B.\ Coluzzi for useful discussions. TSG is fellow of the
Consejo Nacional de Investigaciones Cient\'\i{}ficas y T\'ecnicas
(CONICET, Argentina).

\end{document}